\documentclass[prl,twocolumn,showpacs,preprintnumbers,amsmath,amssymb]{revtex4}
\usepackage{graphicx}
\usepackage{dcolumn}
\usepackage{bm}
\begin{document}

\title{Looking for the Kondo cloud}

\author{J. Simonin}
\affiliation{Centro At\'{o}mico Bariloche, Comisi\'{o}n Nacional de Energ\'{i}a At\'{o}mica,\\
8400 S.C. de Bariloche, R\'{i}o Negro,  Argentina}
\date{agosto 2007}

\begin{abstract}
We study the spatial distribution of the Kondo screening cloud. We find that in 3 \textrm{D} the spatial range of the cloud is limited to distances of the order of the Fermi wavelength due to angular dispersion effects. In lower dimensions this effect is less important and thus the Kondo length determines the localization of the cloud around the magnetic impurity. We clarify the role of the Kondo cloud in the impurity-impurity interactions: the hole coherence factor of the Kondo-doublet interaction is proportional to the amplitude of the Kondo cloud. We single out an experimental setup in which the Kondo cloud could be directly measured.

\end{abstract}
\maketitle

\noindent \textit{Introduction}- When a metallic sample with magnetic impurities is cooled below a certain energy an electronic correlated state develops, screening the impurity spin. That is the Kondo effect, probably one of the most well-studied phenomena in condensed matter physics\cite{soren}, both  experimentally and theoretically. This effect is described by the Anderson Impurity Hamiltonian, whose properties in the different regimes have been studied by a variety of methods, ranging from Wilson's numerical renormalization group\cite{wilson} to ``exact" Bethe ansatz results\cite{bethe}. Most of those results are covered in the book by Hewson, ``\textit{The Kondo Problem to Heavy Fermions}"\cite{hewson}.

Despite all this remarkable research effort there is a basic point that has eluded a clear theoretical description and experimental detection\cite{hand, pascal4}. It is the structure of the Kondo screening cloud and the role of the so-called Kondo length ($\xi_K$). 

The Kondo effect is characterized by a narrow resonance of width $\delta_K$, pinned to the Fermi energy $E_F$. This resonance is intimately related to the formation of a many-body singlet state, comprised of the impurity states and a cloud formed from the metal band states that screens the impurity spin. Its spatial extent is vital for the coupling between neighboring Kondo impurities\cite{hand}. Given the width and localization of the resonance it is straightforward to associate the length $\xi_K = \lambda_F E_F/\delta_K \sim \hbar v_F/\delta_k$ to it, where $\lambda_F$ ($v_F$) is the Fermi wavelength (velocity). Although $\xi_K$ can reach almost macroscopic values given that $E_F/\delta_K \simeq 1000$, it has never been observed in experiments.

The problem possed by such a large value for $\xi_K$ is rather disconcerting \cite{hewson, coleman, varmahf, soren, lobos}. Even a very dilute system with few parts per millon of impurities has a typical inter-impurity separation of about $100$ lattice spacings, much smaller than $\xi_K$. Therefore each impurity has many other impurities inside its screening cloud and cooperative effects must arise. Nonetheless, the impurity resistivity, susceptibility, etc., are observed to be linear in impurity concentration up to very high concentrations, and those quantities fit the theoretical single impurity expectations. 

In addition to the ``classical"  impurities in the 3 \textrm{D} metal  case, the Kondo effect has proven to be a very robust and general phenomenon, arising  in  2 \textrm{D}, adatoms on metal surfaces\cite{adatoms} and quantum corral\cite{alobos} systems, and also in 1 \textrm{D} quantum dots-semiconductor heterostructures\cite{craig}. In all these systems, and given the possibility to use Kondo quantum dots as quantum bits, it is thus relevant to clarify the structure of the Kondo cloud, its role in the interaction between neighboring impurities, and the role of $\xi_K$
\cite{gruner, soren, pascal2, pascal4}.

In the following we solve this long standing puzzle.

\noindent \textit{Theory.} The Kondo effect physics is contained in the Anderson Impurity Hamiltonian. It is composed of a localized orbital (the impurity) of energy $-E_d$, measured from the Fermi level, with a strong internal Coulomb repulsion  $U$, such that the energy cost to put a second electron in it ($-E_d+U$) is, in the Kondo regime, well above $E_F$. Thus double occupance of the impurity is effectively forbidden. This orbital interacts with a band of $N$ extended states, characterized by a half-band width $D$, through an hybridization $V$, that allows an electron to be interchanged between the band and the localized orbital. In the Kondo regime $U \gg E_d,D \gg \delta_K$ the relevant Hamiltonian parameter is the effective Kondo coupling $J = V^2/E_d$ and the Kondo energy is given by $\delta_K = D\ \exp{(-1/2 J_n)}$, where  $J_n= \rho_o J$, $\rho_o$ being the density of band states at the Fermi level.

To analyze the Kondo cloud we use the Kondo singlet variational wave function (VWF) designed by Varma and Yafet\cite{varma} . This method was later used by Gunnarsson and Sch\"{o}nhammer\cite{gunna} to successfully explain the spectra of rare earth compounds, and has become known as the Variational $1/N_S$ Expansion, Ref.\cite{hewson}p.223, where $N_S$ is the degeneracy of the localized orbital, equal to $2$ in the spin one half impurity  case analyzed here. Actually, the expansion parameter can be shown to be $J_n/N_S$. Typical values of $J_n$ are $0.1$, for $\delta_K \simeq 0.01 D$, to  $0.05$, for which $\delta_K \simeq 0.0001 D$. The results of this method are exact in the $J_n/N_S \rightarrow 0$ limit. The Kondo singlet VWF is given by
\begin{equation}\label{sk}
|S_K \rangle = |F\rangle + \sum_{k \sigma} Z_k  \ b^\dag_{k \sigma} d^\dag_{\overline{\sigma}} |F\rangle + \sum_{k q \sigma} Y_{k q}  \ b^\dag_{k \sigma} c^\dag_{q \overline{\sigma}} |F\rangle \ ,
\end{equation}
where $|F\rangle$ is the Fermi sea, \textit{i.e.} the band filled up to the Fermi level. The second quantization operator $b^\dag_{k \sigma}$ creates a hole in the band, \textit{i.e.} it removes an electron from the $k \overline{\sigma}$ band-state below the Fermi level, $ c^\dag_{q \sigma}$ puts an electron in the band above the Fermi level and $d^\dag_\sigma$ puts an electron in the impurity. The $k$ $(q)$ sums are over hole (electron) excitations and $\sigma$ is the spin index. A figure showing these configurations can be seen in Ref.\cite{hewson}. The variational amplitudes of the VWF are $Z_k = \textbf{v} /(\delta_K+e_k)$, and $Y_{k q}= \textbf{v} \ Z_k/(-E_S+e_k+e_q)$, $e_k$ $(e_q)$ being the energy of the hole (electron) excitation in the band and $\textbf{v}=V/\sqrt{N}$. The energy of the Kondo singlet is given by 
\begin{equation}\label{es}
E_S=-E_d  -\delta_K + E_I \ ,
\end{equation}
where the Kondo energy $\delta_K$ comes from the resonance between the $Z_k$ configurations, which use the Fermi sea configuration as a nearly virtual bridge. $E_I$ is the single-body impurity correction, equal to $-J/2$ for the level of approximation used in Eq.(\ref{sk}). It comes from the interplay between each $Z_{k_o}$ configuration and their derived $Y_{k_o q}$ configurations. Therefore Eq.(\ref{es}) ilustrates the versatility of this variational method: it gives both the non-perturbative\cite{kittel} Kondo term as well as the standard perturbative ones. 

\begin{figure}[h]
\includegraphics[width=\columnwidth]{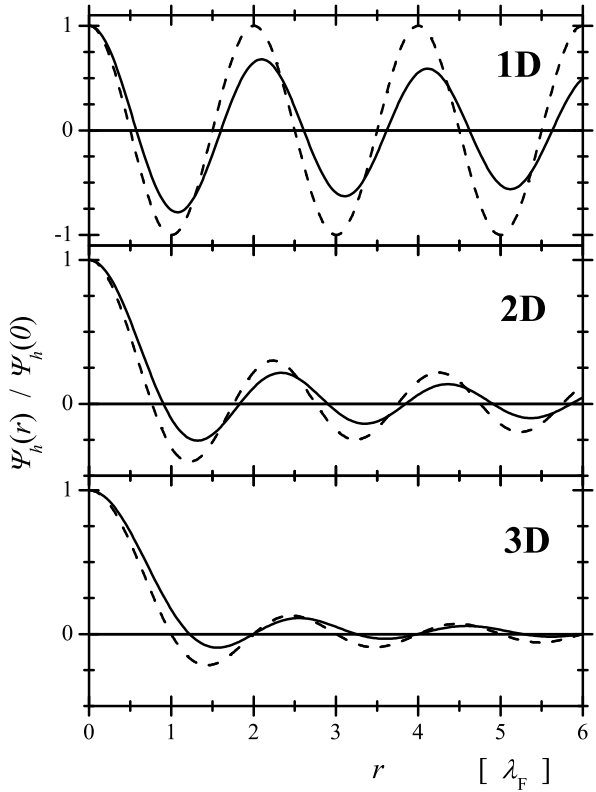}
\caption{Kondo hole orbital profile for $\delta_K = 0.001 D$ in $1$, $2$, and $3$ dimensions. The dashed lines correspond to the respective angular integrals  $\cos(k r)$, $J_0(k r)$ and $\sin(k r)/(k r )$ evaluated at $k = k_F$.  }\label{fig1}
\end{figure}

The square of the norm of $|S_K \rangle$ is given by
\begin{equation}\label{nr}
n_K^2= 1 + 2 \sum_{k} Z_k^2  + 2 \sum_{k q} Y_{k q}^2 \ ,
\end{equation}
and direct evaluation gives the relative total weight of the different set of configurations in the Kondo limit
\begin{equation}\label{wz}
2 \sum_{k} Z_k^2 = 2 \rho_0 V^2 \int_0^D \frac{de_k}{(\delta_K+e_k)^2}\simeq \frac{2\rho_0 V^2}{\delta_K} \gg 1 \ ,
\end{equation} 
and
\begin{equation}\label{wy}
2 \sum_{k q} Y_{k q}^2 \simeq \frac{1}{N}\sum_q (\frac{ V}{E_d})^2 \ 2 \sum_{k} Z_k^2 \ll 2 \sum_{k} Z_k^2 \ ,
\end{equation} 
where the usual $(-E_S+e_k+e_q) \simeq E_d$ approximation has been used in the denominator of $Y_{k q}$. Thus practically all the weight of the VWF is on the $Z_k$ configurations, more precisely in the ones corresponding to holes near the Fermi level. One half of the weight comes from hole states between $\delta_K$ of the Fermi level. To analyze the spatial distribution of the resonant hole we compute the band hole density operator, 
\begin{equation}\label{rhop}
\widehat{\rho}_h(\textbf{r})=\frac{1}{N}\sum_{k k' \sigma}b^\dag_{k \sigma}b_{k' \sigma} \exp{(i(\textbf{k}-\textbf{k}').\textbf{r})} \ ,
\end{equation}
for the singlet, obtaining  
\begin{equation} \label{kcloud}
\rho_K^b(r)=\frac {1}{n_K^2 N} \sum_{k k'}( 1 + \sum_q \frac {\textbf{v}^2}{E_d^2})\ e^{i(\textbf{k}-\textbf{k'}).\textbf{r}} Z_{k} Z_{k'} \ .
\end{equation}
Eq.(\ref{kcloud}) can be rewritten as  $\rho_K^b(r)=C_K(\delta_K) | \Psi_h(r)|^2$ where
\begin{equation}
\Psi_h(r)=\frac{1}{\sqrt{N}} \sum_{k} \ e^{i\textbf{k.r}} Z_{k}=\frac{1}{\sqrt{N}}\sum_{e_k}Z_{e_k} \sum_{\theta_k}e^{i\textbf{k.r}} \ ,
\end{equation}
is the resonant hole wave function\cite{notefi}, and $C_K$ comprise the $r$-independent normalization factors. Given that $Z_k$ depends only on the energy of the hole excitation we separate the sum over $\textbf{k}$ in its angular and modulus parts. The angular integral determines the principal dependence of  $\Psi_h$ on $r$ and on the spatial dimension \textrm{D} of the system.  It gives $\cos{k r}$, the  order 0 Bessel function $J_0(k r)$, and $\sin{k r}/ k r$, in $1$, $2$, and $3$ dimensions respectively (\textit{i.e.} $ \sim \cos{(k r - \pi \alpha /2)}/ (k r)^\alpha $ where $ \alpha = (\textrm{D}-1)/2 \  $). This behavior corresponds to that of radiation waves, and, although each $k$ component is an extended state, it already provides a decrement of the hole cloud amplitude $\sim r^{- (\textrm{D}-1)/2}$. The dashed lines in Fig.\ref{fig1} correspond to those angular integrals evaluated at $k = k_F$.

In Fig.\ref{fig1} there are also plotted (full lines) the results of the whole sum, \textit{i.e.} $\Psi_h(r)$. These sums can be analytically done in $1$ and $3$ dimensions in terms of Cos/Sin-integral functions. The interference between the different terms of this last sum in $|k|$ generates true localization of the hole around the impurity, localization ruled by $\xi_K$ given that $Z_k \sim 1/(\delta_K+e_k)$. But, from a practical point of view, the amplitude of the Kondo cloud decays, in $2$ and $3$ dimensions, in the order of a few times $\lambda_F$ due to its angular dispersion. For $\delta_K \rightarrow 0$, which corresponds to a loosely bound hole with $k \simeq k_F$ , the $\Psi_h(r)$ profile tends to that of the corresponding radiation wave, the dashed lines in Fig.\ref{fig1}. 

The spatial extension of the Kondo cloud opens the possibility for the hole bounded to a given impurity to be captured by an other impurity located at a distance such that the orbital amplitude is still significant. Indeed, this is the case, as that the hole coherence factor of the Kondo doublet interaction\cite{jsfull} is proportional to $\Psi_h(r)$. This connection between the single impurity Kondo cloud and the recently discovered two-impurity Kondo-doublet interaction is not by chance. The Kondo-doublet interaction is mediated by the interchange of a hole between the impurities, \textit{i.e.} the Kondo-doublet is formed by a hole which is partially screening two impurities.

As stated in the introduction the Kondo effect has become relevant in many situations, each one with its own characteristics. 

In 3 \textrm{D} the angular dispersion effectively reduce the extension of the Kondo cloud to few times $\lambda_F$, and $\xi_K$ does not play a significative role, in line with the observed experimental facts. Nevertheless,  plots of $r^x |\Psi_h(r)|^2$, with $x= 2, 3$, show that the hole density decays like $1/r^2$ up to very large values of $r$\cite{d3}. The charge density around a 3 \textrm{D} Kondo impurity has been previously evaluated in Ref.\cite{hewson},p.291 and \cite{sokc}. Those calculations use mean field and/or perturbative solutions of the Anderson Impurity  Hamiltonian and thus fails to capture the non-perturbative Kondo cloud. They found a $1/r^3$ dependence of the charge density that corresponds to the Friedel like oscillations, $\delta_K$ independent, induced by the presence of a normal impurity in a metal. In Ref.\cite{hewson}  the higher order terms ($\sim 1/r^4 $) are attributed to the Kondo effect.

In 2 \textrm{D} the angular dispersion effects are lower than in 3 \textrm{D} and thus the Kondo cloud is effectively extended over a greater region. 2 \textrm{D} Kondo systems has been realized by the deposition of magnetic atoms over metallic surfaces with a strong density of surface states\cite{adatoms}. This situation is particulary interesting because electron density oscillations can be straightforward measured by means of scanning tunnelling microscopy (STM), see Fig.1 in Ref.\cite{knorr} for Co adatoms on a Cu(111) surface. In this setup there is a superposition of the Kondo cloud density oscillations and the Friedel screening charge oscillations\cite{ando}. But they can be discriminated by their $r$-dependence. The Freidel oscillations are given by   
\begin{equation}\label{rof}
\Delta \rho_F(r) \sim \cos{(2 k_F r+\phi_\textsc{D})}/ r^\textrm{D} \ ,
\end{equation}   
they not depend on $\delta_K$, and they normalize to minus the excess charge of the impurity, whereas that the Kondo cloud normalizes to one hole.

\begin{figure}[h]
\includegraphics[width=\columnwidth]{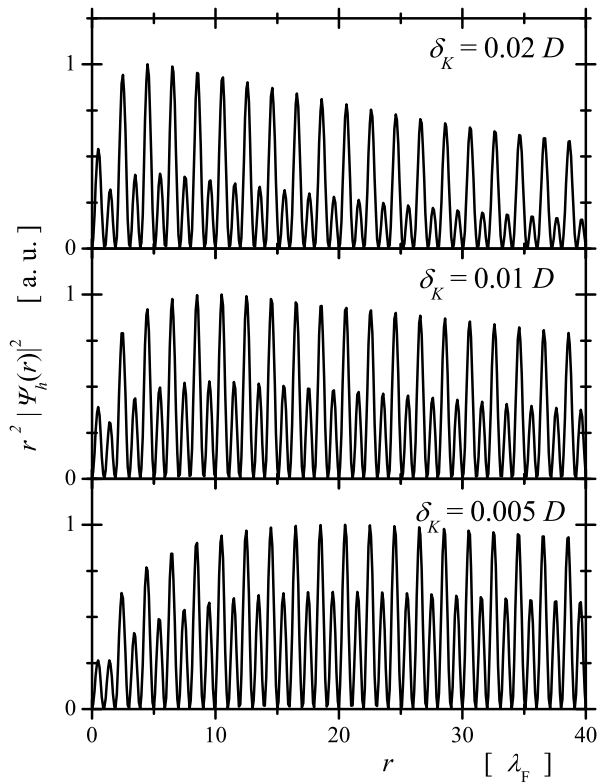}
\caption{2 \textrm{D} Kondo cloud density profile (times $r^2$) for different values of $\delta_K$. Its behavior is quite different from that of the charge Freidel oscillations (a uniform sinusoidal in this representation). The lower $\delta_K$, larger $\xi_K$, the density profile keeps decaying at a rate lower than $1/r^2$ for longer distances.}\label{fig2}

\end{figure}

In Fig.\ref{fig2} we plot $r^2 |\Psi_h(r)|^2$ for 2 \textrm{D} and different values of $\delta_K$. The strong dependence of the Kondo cloud density with $\delta_K$ can be seen. The charge Freidel oscillations (times $r^2$) are a uniform sinusoidal function in 2 \textrm{D}. Instead, in Fig.\ref{fig2}, for the Kondo cloud the decay rate is lower than $1/r^2$ for a distance that scales with $\xi_K$. In fact, the maxima of $r^2 |\Psi_h(r)|^2$ are at $r/\lambda_F \simeq 5$, $10$, and $20$ for the cases plotted in Fig.\ref{fig2}.

Another way to discriminate\cite{gruner} between Friedel and Kondo contributions to the density oscillations is the fact that Kondo cloud disappears above the Kondo temperature. Thus a comparative analysis above and below that temperature will suffice to obtain the Kondo contribution. 

In 1 \textrm{D} there are no angular dispersion effects, and thus the Kondo cloud has an appreciable amplitude over long distances, of the order of $\xi_K $. Given that in these systems formed by quantum dots/wires in semiconductors heterostructures $\lambda_F$ itself is very large, the door is open for the fabrication of very large integrated Kondo circuits. The Kondo-doublet interaction in them can be manipulated by means of tuneable spin-orbit couplings\cite{jsqdqw} and the spin-dot state measured by techniques like the one proposed in Ref.\cite{vk}.     

\noindent \textit{Higher order terms -} The next configurations in the Kondo singlet VWF correspond, from the $Y_{k q}$ ones, to promote a new electron into the impurity, thus generating a new hole in the band (\textit{i.e.} $ \ b^\dag_{k' \sigma'}  b^\dag_{k \sigma} c^\dag_{q \overline{\sigma}}  d^\dag_{ \overline{\sigma}'}|F\rangle \ $). These configurations are equivalent to a $Z_k$ configuration plus a $k'q$ hole-electron pair in the band. They contribute to the total energy of the singlet with a term of order $\ \rho_o J^2 (= J_n J)$. This is the order of the two impurity RKKY interaction\cite{kittel}. In fact the RKKY interaction is generated by the annihilation of the $eh$ pair at a nearby impurity\cite{falicov,jsfull}. The wave-function amplitude of this $eh$ pair goes like $1/r^\textrm{D}$, and its contribution to the density oscillations die off very quickly. The more excitations are present in a given term, more quickly it decays.

\noindent \textit{Conclusions-} We have found the Kondo cloud. Our analysis of its spatial profile shows that in 3 \textrm{D} its range is effectively reduced to a few times $\lambda_F$ due to the angular distribution/dispersion of the orbital. The role of $\xi_K$ in determining the localization of the screening cloud around the Kondo impurity becomes more relevant in 2 and 1 \textrm{D}. We propose that the Kondo cloud can be directly measured in a 2 \textrm{D} adatom setup, Kondo mirage effects have already been measured in a related geometry\cite{mano}.

We also make clear the relation between the spatial extension of the Kondo cloud and the interaction with a nearby impurity. The amplitude of the cloud determines the possibility of the resonant hole to be captured by a neighbor impurity, and thus the hole coherence factor of the Kondo-doublet interaction is proportional to that amplitude.

Beside the experimental possibilities opened by the knowledge of the spatial distribution of the Kondo cloud, further theoretical work is also in order. The Bethe ansatz solution starts from the proposal of an explicit wave function for the Kondo singlet, thus the details of the spatial distribution of the Kondo cloud could be also extracted from the ansatz.   

We thank CONICET (Argentina) for partial financial support.

\bibliography{cloud}

\begin{thebibliography}{27}
\bibitem[{not()}]{notefi}
\bibinfo{note}{One can evalute directly the hole field operator.}
\bibitem[{d3()}]{d3}
\bibinfo{note}{With the flat band approximation, which overstimate the weight
  of long wavelenght hole states, this is also the asymptotic behavior. A more
  realistic density of states correct this point.}
\end{thebibliography}

\end{document}